# FRAMEWORK FOR MANAGING CYBERCRIME RISKS IN NIGERIAN UNIVERSITIES

Bukhari Badamasi, American University of Nigeria, bukhari.badamasi@aun.edu.ng

Samuel C. Avemaria Utulu, American University of Nigeria, samuel.otulu@aun.edu.ng

**Abstract:** Universities in developing countries, including those in Nigeria, experience cybercrime risks due to poor management of their cyber spaces and resources. The outcome of these cybercrimes are threats and breaches of universities' cyber security. The threats and breaches have resulted in substantial financial, social, and intellectual property losses. In the recent past, Nigerian universities have started to respond to these cyber-attacks. Many of them now invest in anti-cybercrime tools and programs to mitigate cyber security threats and breaches. Despite this, the number of times Nigerian universities suffer from cyber-attacks and the losses that result from them keeps increasing. Our observation, however, indicates that most Nigerian universities run their cyber security without using scientifically derived frameworks that spell out how to manage threats and breaches that emanate from within and outside them. We consider this a problem to ongoing efforts made by Nigerian universities to mitigate cyber security threats and breaches. The study reported in this paper was therefore, carried out to explicate how Nigerian universities can develop actionable frameworks that can help them to mitigate cyber security threats and breaches. The study is based on literature review and propose how an actionable framework that Nigerian Universities can adopt to setoff cybersecurity programs can be developed. The process comprises of problem identification, description of objectives, designing and developing the artefact, testing, and evaluating the artefact, and communicating the result. We conclude that the framework provides a lucrative starting point for Nigerian universities to setoff efficient and effective cyber security program.

## 1. INTRODUCTION

Universities are highly dependent on the Internet and cyberspaces within it to actualize their statutory responsibilities. According to Li et al. (2018), cyberspace refers to an interconnected and interdependent information technology (IT) based networks that include the Internet, telecommunication networks, computer systems, and social systems. Aheleroff et al. (2021; p.5) also stated that "cyberspace has emerged as a powerful interconnected digital technology with the ability to achieve the most complex manufacturing paradigms due to the advancement features associated with Big Data, Internet of Thing, and Blockchain technology". Cyberspaces have evolved tremendously over the years, providing a variety of digital platforms universities need to manage teaching, learning, research, community development and administration (Taylor, 2017). Cyberspaces have helped universities to manage admission processes, students' life issues, finance, examinations and records, and to facilitate academic processes (Hunton, 2011). It follows that every contemporary university carries out its statutory responsibilities and provide services to staff, students, parents and guardians, funding agencies, government, accreditation agencies, and other stakeholders using cyber spaces. Despite the advantages of cyberspaces offer universities, they also pose critical threats and challenges to their well-being and operations. This is given the evolution of cybercrimes and their growth in the recent past. Reports in the extant literature show that a variety of culprits that have become very challenging to identify and apprehend have emerged within cyberspaces universities use to actualize their statutory responsibilities.

Recent incidences of cybercrimes in universities has resulted to a new understanding among cybersecurity scholars that cybercrime is not limited to financial and related institutions. Demers et al. (2017) for instance, revealed that cybercrimes perpetrated against universities is increasing at an alarming rate. Demers and his colleagues argued that the education sector is second in the league of industries that suffer cybersecurity threats and breaches. According to France-Presse (2020), the FBI and the Cybersecurity and Infrastructure Security Agency (CISA) both announced that organizations researching COVID-19, particularly universities, were at risk given China's attempt to steal coronavirus research data. Attempts were made by Chinese government-affiliated organizations and others to unlawfully acquire valuable intellectual property and public health data related to vaccines, treatments, and testing (Xie, 2020). Walker (2020) also reported that the United Kingdom (UK) Security Minister alleged that he was more than ninety five percent certain that state-sponsored hackers backed by Russian Government targeted organizations and universities in the UK and Canadian that were working on a coronavirus vaccine. There are similar reports in the extant literature indicating that the National Cyber Security Agency (NCSA) in Nigeria and its counterparts in other countries were adamant that the attacks on drug companies and research groups were carried out by "Russian Intelligence Agency" (Parsons, 2020). Sobers (2021) further stated that universities were ranked as the most dangerous place for an individual to reveal sensitive information. Opinions propagated by Demers et al. (2017) and Sobers (2021) fits into recent occurrences of cybercrime experiences of Nigerian universities.

Examples of cybercrime cases in universities in Nigeria include the incidence of Denial of Service (DoS) attack in which an unknown person abused the Network Time Protocol (NTP) server in the Federal University of Technology Akure (Mojeed, 2020). Madonna University in the Eastern Nigeria also reported that hackers accessed and tempered with over 25,000 data in their database (Egbunike, 2019). In 2016, 2017, and 2018, Ahmadu Bello University, Zaria, Nigeria disclosed incidents that potentially exposed a serious breach in the university website that resulted in a compromise of students and staff data (Bukhari, 2018). In another incident, a university staff was found compromising the admission records and printing fake admission letters for some applicants. In a similar incident, a staff of the Management Information System (MIS) Unit was apprehended for breaching the security protocols and conniving with some students of the University to illegally allocate rooms and print admission letters (Bukhari, 2018). All these challenges are good examples of cyber security issues Nigerian universities contend with.

Studies have however, shown that to manage cyber security risks, there is the need to plan for them (Clausen, 2019; Mamogale, 2011). Planning for cyber security risks requires putting in place adequate measures to avoid cyber-attacks and/or to manage their effects if they eventually occur (Alpert, 2012; S. T. Clausen, 2019; Kuusikallio, 2017; Mamogale, 2011). It is therefore, follows that, that major challenge confronting Nigerian universities is the need to develop frameworks that will serve as guides to programs they develop to avert cybercrimes and to reduce the effects of cybercrimes when they are perpetrated. In developed countries, a good number of universities, have cyber risk management frameworks, this however, cannot be said about Nigerian universities and most universities in developing countries (Singh & Joshi, 2017). The case in Nigeria, with regards to the small number of universities that have frameworks for managing cybercrimes, is unique. Ryder and Madhavan (2019) opinion that strategies used to fight cybercrime in developing countries are weighted towards short-term responses and IT challenges and that the strategies do not always spell out how to manage the consequences of cybercrime incidences succinctly captures the situations in Nigeria.

In our opinion, a good cybercrime management framework should be weighted towards long-terms responses and should provide grounds for the careful monitoring of cyberspace before, during and after cybercrime occurs. The indication is that a cybercrime management framework must be holistic and must bring into bear every aspects of cyberspace management. This is given that cybercrime management involves different tasks and skills and diverse stages. Our study is an ongoing large scale and longitudinal study that is being carried out in the Nigerian university context. The study is relevant because Nigerian universities are among universities in developing country contexts with high risk of cybercrime incidents (Eboibi, 2020). Consequently, the research study reported in this paper is a part of the large scale and longitudinal study. This particularly study was informed by the question, how perspectives in the design science method can facilitate the development of cybersecurity management framework for Nigerian universities. To answer this question, five specific questions derived from the design science method process were raised including: what are the cybersecurity problems Nigerian universities are facing and what are the likely problems they will face in the future? What should the objectives of cybersecurity programs of Nigeria universities be? How can appropriate cybersecurity program be designed and implemented by Nigerian universities? How can the appropriateness and adequacy of Nigerian universities' cybersecurity programs be tested and evaluated? How can cybersecurity programs of Nigerian universities be communicated to necessary stakeholders? The paper aims to develop a cybersecurity management framework that is useful to Nigerian universities and universities in other developing countries that operate in socio-technical environments that are similar to those in the Nigerian university system. The remaining part of the paper includes review of related literature, methodology, proposed framework, and conclusion and limitations.

## 2. REVIEW OF RELATED LITERATURE

### 2.1 Cybercrime

Cybercrime has become the world's second-largest man-made risk (Soomro & Hussain, 2019). It encompasses all illegal activities perpetrated by scammers, hackers, and internet fraudsters. The illegal activities may include human activities carried out to gain illegal access to data and information or sending spam, malware, worms into devices, networks, or even organizational information systems and global connection to make them malfunction (Mary, 2016). Cybercrimes have had debilitating effects on individuals, governments, organizations, and universities (Adesina & Ingirige, 2019). It has cost billions of dollars' worth of damages, data loss, and website defacement. It has sent many governments, organizations, and individuals into bankruptcy and global shock (De Paoli et al., 2020). Cybercrime has remained a major threat to universities, particularly as it touches that core mandates of teaching, learning, research, community services, and administration and management of staff and student records (Bukhari, 2018). There are indications in the extant literature that the COVID-19 era is likely to result in a surge in the number of cybercrimes perpetrated against universities. Traxler et al. (2020) for instance, opined that COVID-19 has forced an increase in the dependence on cyberspace for both individuals and organizations and that this is likely to result in cybersecurity threats and risks. Morgan (2020) also observed that more than four thousand malicious COVID-19 related websites appeared on the internet within months of the first COVID-19 pandemic in 2020. He argued that the occurrence of cybercrime in 2021 would be in every eleven seconds; a fit that is almost four times the average in 2020 (every nineteen seconds) and almost twice the pace in 2019 (every forty seconds) (Morgan, 2020). Meanwhile, cybercrime is projected to cost the global economy $6 trillion yearly from 2021, up from $3 trillion in 2015. While from 2025, cybercrime will cost $10.5 trillion annually. Besides the United States and China, cybercrime would eventually be the world's third-largest economy (Sausalito, 2020). Losses to ransomware are projected to cost the world $20 billion by 2021, which is 57 times what they were in 2015 ($325 million) (Chapman, 2019; Morgan, 2020). As a result, ransomware is the fastest-growing form of cybercrime even in universities. Furthermore, spear-phishing emails are used in 91 percent of cyberattacks, infecting organizations and universities.

There is no doubt that insights in the extant literature indicate the need for a robust cybersecurity management framework that can provide actionable knowledge to organizations and universities. In its entirety, cybercrime risk management frameworks provide a multiplicity of guiding principles and action plans aimed at addressing cybercrime and its related incidents. In Nigeria, the Office of the National Security Adviser (NSA) in collaboration with the National Information and Technology Development Agency (NITDA) shares similar views with Microsoft and the National Institutes of Standards and Technology on the threats to be addressed through the implementation of a consolidated cybersecurity framework (Osho & Onoja, 2015). Although, this is mainly for organizations like the financial sector, oil and gas, and other conglomerates. This situation makes our attempt at producing an actionable cybersecurity framework for Nigerian universities a worthwhile venture.

## 2.2　Cybersecurity issues in Universities

Cybersecurity is derived from two words: cyber and security. According to Valeriano and Maness (2015), cyber is related to the technology which contains systems, network and programs or data (Valeriano & Maness, 2015). On the other hand, Schneier (2009) argued that security relates to protection of systems, network, application, and information. Sanoo (2018) further described cybersecurity as the protection of interconnected systems, including hardware, software, and data from cyber-attacks. One important factor that Valeriano and Maness (2015), Sanoo (2018) and Schneier (2009) did not include in their definitions is the social: values, norms and cultures conjectured by organizations as suitable and logical ways of behaving and relating with others. Consequently, we take cybersecurity to be primarily about people and the social structures and work processes they create, and technologies they put together to encompass the full range of threat reduction, vulnerability reduction, deterrence, international engagement, incident response, resiliency, and recovery policies and activities, including computer network operations, information assurance, law enforcement. Some authors have proposed similar view about cybersecurity in the extant literature. These authors argued that the body of technologies, processes, and practices designed to protect networks, devices, programs, and data from attack, theft, damage, modification or unauthorized access constitute cybersecurity management protocols (Abu-Taieh, 2017; Rashid et al., 2018; Sanoo, 2018). Cybersecurity is used in different areas in universities, and each university has its area of usage or application. Most universities in the world has deployed cybersecurity to protect their network, and data on the cyberspace. Universities in the developed countries such as University of Arizona, University of Edinburgh, University of Bristol, University of Sheffield, Princeton University, University of Illinois, University of Leicester, Carnegie Mellon University, and University of Pittsburg uses different cybersecurity frameworks to manage cybersecurity risk. The framework deployed helps to facilitate the strategic vision of the universities and facilitate the protection of information systems against compromise of its confidentiality, integrity, and availability (Webb & Hume, 2018). Whilst doing this, it recognizes the ability to discover, develop, and share knowledge among employees. While Nigerian universities are reluctant about cybersecurity framework.

Several studies were conducted to examine issues concerning cyber security threats in Nigerian Universities. A study conducted by Ekpoh et al. (2020) in which factors that served as cybersecurity threats to universities were examined at University of Lagos, it was found in the study that there existed a strong positive relationship between location, culture, facilities and personnel security of universities, while a weak, positive correlation existed between school climate and personnel security. The study concluded that indiscipline, poor staff and student's safety and security awareness, inadequate capacity building for security personnel, poor funding of institutions, and outdated security framework, were the major determinants of security lapses in Nigerian Universities. A study was conducted by Dagogo (2005) on the role of security agents in curbing cybercrimes in Universities in the North-East Nigeria using seven tertiary institutions. The study revealed that training and re-training of security personnel and cybersecurity expert significantly affect their level of service delivery. Statistically, Nigerian universities ranked 43 in Europe, the Middle East and Africa and ranked third among the nations that commit cybercrime in the world (Makeri, 2017).

## 2.3  Cyber Security Framework

Cyber security management framework has to do with mitigating cybersecurity risks. It is a relatively new and growing aspect of risk management in organizations. Organizations, including universities, face risks due to natural occurrences, human resource failures, third-party contractors, financial mayhem, chaotic conditions, and security breaches. Risk is an unpredictable occurrence of an incident or situation in organizations or universities that has negative impact, such as time, cost, or quality (Mikkola et al., 2020). A risk may have one or more triggers or causes, and if it happens, it may have diverse effects. Every facet of universities' information systems and technical and social environments can face diverse risks which are likely to be caused by poor preparation and management procedures, and lack of centralized management systems (Hollis, 2015). Consequently, risk management (RM) has become an important and integral part of cybersecurity management in organizations over the last few decades (Whitehead, 2020). It encompasses the mechanisms involved in hazard preparation, assessment, interpretation, reactions, and threat management and regulation (Purohit et al., 2018), which is further described as a role that enables and adds value to organizations while increasing the likelihood of achieving strategic objectives. Meanwhile, every organization that wants to excel, must develop strong capabilities to handle complex risks. Contemporary organizations must build an empowering atmosphere that reduces the negative consequences of risk. The idea that it is advisable that universities have cybersecurity frameworks stems from insights propagated in the RM domain. this is given the relationship between critical aspects of RM and those of cybersecurity frameworks namely, techniques, processes, and resources used to define and manage risks (Aven & Renn, 2010).

There are several cybersecurity frameworks in use across the world, depending on person or organizational preference and adaptation (Pattinson et al., 2018). Defense in Depth and Defense in Breath, NIST Cybersecurity Framework, The Lockheed Martin Kill Chain, Specified Frameworks, Global Cybersecurity Index, and Cybersecurity Risk Framework (Smith, 2019). However, the most widely used framework developed by the US National Institute of Standards and Technology (NIST), offers a high-level taxonomy of cybersecurity outcomes as well as a methodology for assessing and managing them. According to NIST (2020), the NIST Cyber Security Framework (CSF) describes five core functions that organizations should address to pro-actively manage cybersecurity threats to their business operations; identification, detection, protection, reaction, and recovery. Another cybersecurity framework widely used in organizations is the Global Cybersecurity Index (GCI) conceptual framework designed by the International Telecommunication Union (ITU) in collaboration with ABI research institutes (ITU, 2015). The framework seeks to evaluate a country or organization's development in cybersecurity systematically. The objective of this framework is for cybersecurity to be a focal point in information systems organizations and users of those systems (ITU, 2015). Five critical factors were identified as the constructs which determines the dimension of cybersecurity within organizations (ITU, 2015; Maarten et al., 2015; Stein, 2008). These constructs include technical measures, legal measures, capacity building, co-operation, and organizational measures.

Although existing frameworks are simple and actionable, most of them do not directly address cybersecurity issues faced by universities. The peculiarities of universities in developing country contexts also raises concerns that are not directly addressed in existing international frameworks. To control cybersecurity threats, universities must have a thorough understanding of their socio-technical contexts, operations, drivers, and security issues. Since the threats, goals, and processes of each university are distinct, the techniques and approaches used to achieve the objectives that inform cybersecurity frameworks usually differ. As a result, this study invites Nigerian universities and universities in similar developing country context to be clear about their socio-technical contexts, operations, drivers, and security issues. To achieve this clarity of purpose, Nigeria universities must provide answers to the questions raised in this study namely: what are the cybersecurity problems Nigerian universities are facing and what are the likely problems they will face in the future? What should the objectives of cybersecurity programs of Nigeria universities be? How can appropriate cybersecurity program be designed and implemented by Nigerian universities? How can the appropriateness and adequacy of Nigerian universities' cybersecurity programs be tested and evaluated? How can cybersecurity programs of Nigerian universities be communicated to necessary stakeholders? Painstaking and scientifically derived answers will provide grounds for a unique cybersecurity framework.

## 3. METHODOLOGY

The study adopted the interpretive philosophy. The Interpretivist philosophy will have scholars believe that social realities, such as cybercrime and cybersecurity, and IS artefacts including frameworks, are socially constructed, that is manmade (Ngwenyama, 2014). It also makes scholars to work with the assumption that the human actors involved in the use of IT and the IT itself are subjective and act based on socially constructed notions (Utulu & Ngwenyama, 2017). The method adopted for the study is the literature review method. Extensive review of the literature was carried out on themes relating to cybercrime, cybersecurity, and design science research approach. The literature review method adopted in the study was more like the snowball technique, where works cited by the works we consider primary to our debate and the works that cited them were selected and used to come up with the arguments that we presented in the paper. This literature review method has been used by Avgerou (2008), Heeks (2017) and Olagunju and Utulu (2021). The literature reviews method is unlike the systematic literature review or grounded theory literature review that are based on premeditated procedures (Okoli & Schabram, 2010; Utulu et al., 2013; Wolfswinkel et al., 2013). Theoretical perspectives of the design science research approach were used to come up with the cybersecurity management framework that we proposed. Baskerville et al. (2018), argue that design science method involves the creation of an artefact, framework, model, or theory in which the current state of practice can be improved together with the existing knowledge. Hevner and Chatterjee (2010) further stated that design science method is a problem-solving paradigm which results, among others, in the development of models or frameworks that are useful to solving practical problems.

## 4. PROPOSED FRAMEWORK

### 4.1 Identifying Cybersecurity Threats Nigerian Universities are likely to face

The primary tasks of universities are production and dissemination of scientific knowledge research and scientific knowledge publication. Globally, universities are vulnerable to diverse forms of cybersecurity problems, including theft of intellectual property, compromise of student and staff records and hacking of university portal (Oliver, 2010). Moreover, different forms of cybercrime ranging from admission falsification, impersonation, illegal room allocation, website defacement, hacking of log-in details, printing of fake admission letters among others are the cybersecurity challenges facing Nigerian universities (Bukhari, 2018; Igba et al., 2018; Okeshola & Adeta, 2013). The emerging problems that Nigerian Universities may face are numerous. Some of the problems include beneficiary of a will scam. According to Bian et al. (2018), will scam occurs when cybercriminal send e-mail to claim that the victim is the named beneficiary in the will of an estranged and stands to inherit and estate worth millions. Another emerging cybersecurity challenge is online charity. In online charity, cybercriminals host websites as if they are charity organizations. They use the websites to solicit for monetary and material donations (Saulawa & Abubakar, 2014). The possibility that cybercriminal can set up fake websites and lure donors to donate to universities is high. Another cybercrime Nigerian Universities suffer from is computer/Internet service time theft. Culprits develop means of connecting privately owned cyber cafes to networks owned by universities in ways that are difficult to detect and thereby run their cafés at the expense of the universities (Oliver, 2010).

## 4.2 Setting Objectives for Nigerian Universities' Cybersecurity Programs

The main objectives of cybersecurity programs are to help Nigerian universities reduce the vulnerability of the information systems and networks they operate in cyberspaces. Setting adequate and appropriate objectives for cybersecurity programs can be a very complex task for universities, including Nigerian universities (Igba et al., 2018). This is because the extent to which cybersecurity programs of Nigerian universities can reach is highly dependent on the objectives the set. Consequently, the values cybersecurity objectives of Nigerian universities are funtions of their understanding of cyber security threats they face, the extent they are able to articulate and share information about the cybersecurity threats with necessary stakeholders and the ways they are able to translate their understanding of the cybersecurity threats to cybersecurity management policies and practices (Bian et al., 2018; Saulawa & Abubakar, 2014). The extent nigerian universities collaboratively work with public, private, and international entities with regards to their cybersecurity programs is also a function of the adequacy and appropriateness of the cybersecurity objectives they set. Cybersecurity objectives provide the framework of reference that helps organizations to understand current trends in cybercrimes and solutions that are effective and efficient in tackling them. Cybersecurity objectives also provide grounds the measure levels of integrity, reliability and efficiency of cybersecurity programs.

## 4.3 Techniques for Managing Cyberattacks by Nigerian Universities

Cybersecurity policy framework is the first point of action for managing cybersecurity threats. Although it evolves from cybersecurity objectives set by universities, it spells out what Nigerian universities should do and how to do what they have to do with regards to cybersecurity threats. Each universities cybersecurity policy should be integrated with other universities and organizations and should provide room for determine other universities and organizations should policy to avert or management cybersecurity attacks (Makeri, 2017). An appropriate cybersecurity framework should also define required cybersecurity education and training universities need to provide members of university communities. It is also necessary to educate members of university communities and various organizations universities deal with in the best practice for effective cybersecurity management. It should also spell out cybersecurity requirements of other organizations the universities deal with. For example, some universities in the developed countries have a policy that all systems in their purview must meet strict security guidelines (Ekpoh et al., 2020). Automated updates are sent to all computers and servers on the internal network, and no new system is allowed online until it conforms to the security policy (Iriqat & Molok, 2019). Cybersecurity management resources required to avert or manage cybersecurity attacks are also to be spelt out in cybersecurity policy frameworks. Cybersecurity policy frameworks Nigerian universities use should also spell out the role ISPs are to play within universities' cyberspace and how to ensure high level of security at servers in order to keep clients secure from all types of cyberattacks (Odinma, 2010).

## 4.4 Cybersecurity Programs Appropriateness and Adequacy Assessment

An important part of the cybersecurity management framework proposed in this paper is making room to assess the appropriateness and adequacy of the entire cybersecurity management framework. This could be done in two way (Pavol Zavarsky & CISM, 2014). First, is appropriateness and adequacy assessment that is based on assumptions (Armenia et al., 2021). Second, is the appropriateness and adequacy assessment that is based on experience (Glantz et al., 2016). The first option occurs given that appropriateness and adequacy are determined before the occurrence of cyberattack. The second option occurs after a cyberattack when a university assesses its cybersecurity management framework vis-à-vis the nature and strategy cyberattack it suffered. The attack may not be a serious attack, but it provides avenue for cybersecurity management framework appropriateness and adequacy assessment. These two approaches to assessing the appropriateness and adequacy of universities' cybersecurity management frameworks can help those concerned to modify existing cybersecurity management frameworks. They help to open avenue for constructive feedbacks from those concerned. The use of stakeholders' feedbacks and recommendations are made possible by cybersecurity appropriateness and adequacy assessment.

## 4.5   Communicating Cyberattack and Management Outcomes to Stakeholders

This requirement is important, and can be used during two different situations. The first situation is pre-cyberattack situation while the second situation is post-cyberattack situation. During the pre-cyberattack situation, universities are expected to communicate how their cybersecurity management framework works and the role of each stakeholder group. In the second situation, universities are to communicate loopholes in the cybersecurity management framework that resulted to the cyberattack experienced and how the updated cybersecurity management framework solves the problems that resulted from the loopholes. Communicating ideas across large organization is a complex task (Smith, 2019). Heide et al. (2018; p. 2) in his article "Expanding the Scope of Strategic Communication: Towards a Holistic Understanding of Organizational Complexity", describes strategic communication as an academic movement that has been formulated as an ambition to break down the silos surrounding closely related communication disciplines and create unifying framework that integrates public relations, organizational communication, marketing communications and other areas" Organizations should communicate strategically to purposefully fulfill their overall missions. This complexity is also applicable to efforts made by organizations to communicate cybersecurity management framework across the length and breadth of organizations. The complex nature of cybersecurity threats and the difficulty in knowing the perpetrators and understanding their motives makes the act of communicating cybersecurity management frameworks across the length and breadth of organizations a complex endeavor. Aside this, some aspects of cybersecurity management frameworks that universities may use may be made clandestine. So, it is important to know and understand those that these aspects should be communicated to and how to effectively and efficiently do this without jeopardizing the overall cybersecurity management program.

## 5. CONCLUSION AND LIMITATION

The rapid expansion of cyberspaces and universities transfer of their major activities and operations into cyberspaces have led to the increase of cybercrime perpetrated against universities. The frequency in which universities across the globe suffers from cybersecurity attacks indicate the need for Nigerian universities to develop cybersecurity management frameworks that they ae use to coordinate their cybersecurity management programs. This is not to say that Nigerian universities do not have cybersecurity strategies they use. It however, indicates that they need to make concerted efforts towards formalizing and documenting their cybersecurity management strategies into actionable frameworks. This paper presents a proposed framework that provides good grounds for Nigerian universities to set off their actions towards developing actionable cybersecurity frameworks. The paper proposes three-stage based cybersecurity management framework for Nigerian universities namely, pre-cybersecurity readiness, cyberattack management and post-cyberattack activities. Each stage was broken down into actionable processes. The limitation of the framework and by extension this paper, is that it is not based on empirical research study. It is based on literature review. An empirical study would have provide empirically derived insights on how the proposed cybersecurity management framework will work in real life situations. However, the paper and the proposed framework provide grounds for conducting empirical studies on cybersecurity management framework for universities.

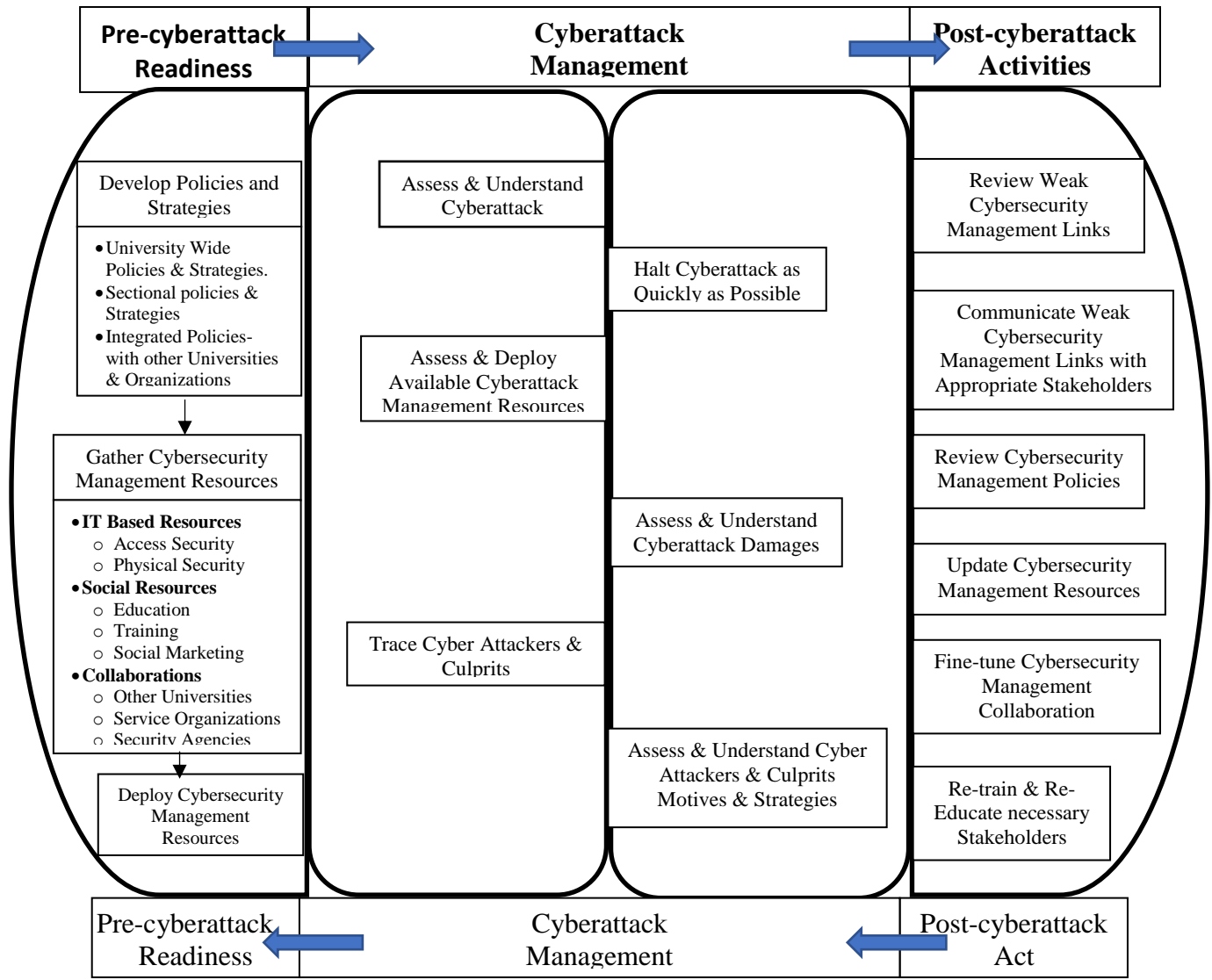

**Figure 1: Nigerian Universities' Cybersecurity Management Framework**

## REFERENCES

Abu-Taieh, E. M. (2017). Cyber security body of knowledge. 2017 IEEE 7th International Symposium on Cloud and Service Computing (SC2),

Adesina, R., & Ingirige, B. (2019). Dismantling barriers to effective disaster management in nigeria. 14th International Postgraduate research conference 2019: Contemporary and Future Directions in the Built Environment,

Aheleroff, S., Xu, X., Zhong, R. Y., & Lu, Y. (2021). Digital twin as a service (DTaaS) in industry 4.0: an architecture reference model. *Advanced Engineering Informatics*, *47*, 101225.

Alpert, B. S. (2012). *College and University Disaster Management: The Impact of Leader Behavior on Response and Recovery from Disaster*

Armenia, S., Angelini, M., Nonino, F., Palombi, G., & Schlitzer, M. F. (2021). A dynamic simulation approach to support the evaluation of cyber risks and security investments in SMEs. *Decision Support Systems*, 113580.

Aven, T., & Renn, O. (2010). Risk management. In *Risk Management and Governance* (pp. 121-158). Springer.

Avgerou, C. (2008). Information systems in developing countries: a critical research review. *Journal of information Technology*, *23*(3), 133-146.

Baskerville, R., Baiyere, A., Gregor, S., Hevner, A., & Rossi, M. (2018). Design science research contributions: finding a balance between artifact and theory. *Journal of the Association for Information Systems*, *19*(5), 3.

Bian, S., Deng, Z., Li, F., Monroe, W., Shi, P., Sun, Z., Wu, W., Wang, S., Wang, W. Y., & Yuan, A. (2018). Icorating: A deep-learning system for scam ico identification. *arXiv preprint arXiv:1803.03670*.

Bukhari, B. (2018). *Effects of Security Protocols on Cybercrme in Ahmadu Bello University, Zaria* [Academic Masters, University of KwaZulu Natal, South Africa].

Chapman, J. (2019). How safe is your data? Cyber-security in higher education. *Higher Education Policy Institute Policy*.

Clausen. (2019). Justifying military intervention: Yemen as a failed state. *Third World Quarterly*, *40*(3), 488-502.

Clausen, S. T. (2019). Enabling the Implementation of Drones into Local Disaster Preparedness Key considerations from challenges and lessons learned in Chile.

De Paoli, S., Johnstone, J., Coull, N., Ferguson, I., Sinclair, G., Tomkins, P., Brown, M., & Martin, R. (2020). A Qualitative Exploratory Study of the Knowledge, Forensic, and Legal Challenges from the Perspective of Police Cybercrime Specialists. *Policing: A Journal of Policy and Practice*.

Demers, G., Harrington, S., Cianci, M., & Green, N. (2017). Protecting Colleges & Universities Against Real Losses in a Virtual World, 33 J. Marshall J. Info. Tech. & Privacy L. 101 (2017). *The John Marshall Journal of Information Technology & Privacy Law*, *33*(2), 3.

Eboibi, F. E. (2020). Concerns of cyber criminality in South Africa, Ghana, Ethiopia and Nigeria: rethinking cybercrime policy implementation and institutional accountability. *Commonwealth Law Bulletin*, *46*(1), 78-109.

Egbunike, N. (2019). *Nigerian students face cybercrime charges for criticising their university online*. https://globalvoices.org/2019/07/11/nigerian-students-face-cybercrime-charges-for-criticising-their-university-online/

Ekpoh, U. I., Edet, A. O., & Ukpong, N. N. (2020). Security Challenges in Universities: Implications for Safe School Environment. *Journal of Educational and Social Research*, *10*(6), 112-112.

France-Presse, A. (2020). *US Says China Trying to Steal COVID-19 Vaccine Research*. https://www.voanews.com/covid-19-pandemic/us-says-china-trying-steal-covid-19-vaccine-research

Glantz, C., Somasundaram, S., Mylrea, M., Underhill, R., & Nicholls, A. (2016). Evaluating the maturity of cybersecurity programs for building control systems. *US Department of Energy Office of Scientific and Technical Information*.

Heeks, R. (2017). Decent work and the digital gig economy: a developing country perspective on employment impacts and standards in online outsourcing, crowdwork, etc. *Development Informatics Working Paper*(71).

Heide, M., von Platen, S., Simonsson, C., & Falkheimer, J. (2018). Expanding the scope of strategic communication: Towards a holistic understanding of organizational complexity. *International Journal of Strategic Communication*, *12*(4), 452-468.

Hevner, A., & Chatterjee, S. (2010). Design science research in information systems. In *Design research in information systems* (pp. 9-22). Springer.

Hollis, S. (2015). The role of regional organizations in disaster risk management. In *The Role of Regional Organizations in Disaster Risk Management* (pp. 1-12). Springer.

Hunton, P. (2011). A rigorous approach to formalising the technical investigation stages of cybercrime and criminality within a UK law enforcement environment. *Digital investigation*, *7*(3-4), 105-113.

Igba, D., Elizabeth, C., & Nwambam, A. S. (2018). Cybercrime among University Undergraduates: Implications on their Academic Achievement. *International Journal of Applied Engineering Research*, *13*(2), 1144-1154.

Iriqat, Y. M., & Molok, N. N. A. (2019). Information security policy perceived compliance among staff in palestine universities: an empirical pilot study. 2019 IEEE Jordan International Joint Conference on Electrical Engineering and Information Technology (JEEIT),

ITU. (2015). *Global cybersecurity index & cyberwellness profiles report* (Cybersecurity, Issue. I. T. Union. https://www.itu.int/pub/D-STR-SECU-2015

Kuusikallio, V. (2017). Community-based disaster preparedness in The Kimbilio Women´s Shelter and Education Center.

Li, F., Li, Z., Han, W., Wu, T., Chen, L., Guo, Y., & Chen, J. (2018). Cyberspace-oriented access control: A cyberspace characteristics-based model and its policies. *IEEE Internet of Things Journal*, *6*(2), 1471-1483.

Maarten, G., Artur, U., Erik, F., & Michel, R. (2015). *A meta-analysis of threats, trends, and responses to cyber attacks* (Assessing Cyber Security, Issue. T. H. C. f. S. Studies. https://hoffmannbv.nl/sites/default/files/Report%20Assessing%20Cyber%20Security%2016%20april%202015.pdf.

Makeri, Y. A. (2017). Cyber Security Issues in Nigeria and Challenges. *International Journal*, *7*(4).

Mamogale, H. (2011). Assessing disaster preparedness of learners and educators in Soshanguve North schools. *Bloemfontein, South Africa: The Disaster Management Training and Education Centre for Africa, the University of the Free State*.

Mary, L. (2016). IT Security and Privacy.

Mikkola, M., Oksanen, A., Kaakinen, M., Miller, B. L., Savolainen, I., Sirola, A., Zych, I., & Paek, H.-J. (2020). Situational and Individual Risk Factors for Cybercrime Victimization in a Cross-national Context. *International Journal of Offender Therapy and Comparative Criminology*, 0306624X20981041.

Mojeed, M. (2020). How Nigerian University Launched Massive Cyberattacks Against Premium Times. https://allafrica.com/stories/202007280025.html

Morgan, S. (2020). Cybercrime To Cost The World $10.5 Trillion Annually By 2025 Cybercrime Magazine. In.

Ngwenyama, O. (2014). Logical foundations of social science research. In *Advances in Research Methods for Information Systems Research* (pp. 7-13). Springer.

NIST. (2020). *CYBERSECURITY FRAMEWORK*. https://www.tenable.com/lp/campaigns/20/whitepapers/adhering-to-the-nist-framework-with-tenable-ot/?utm_campaign=gs-{9662775243}-{100779850978}-{426501511627}_00021238_fy21q1&utm_promoter=tenable-indegy-nb-00021238&utm_source=google&utm_term=%2Bnist%20%2Bframework&utm_medium=cpc&utm_geo=emea&gclid=EAIaIQobChMIjbXbsunm7wIVAtWyCh2j7g6IEAAYASAAEgIu0PD_BwE

Odinma, A. (2010). Cybercrime & Cert: Issues & Probable Policies for Nigeria. *DBI Presentation, Nov*, 1-2.

Okeshola, F. B., & Adeta, A. K. (2013). The nature, causes and consequences of cyber crime in tertiary institutions in Zaria-Kaduna state, Nigeria. *American International Journal of Contemporary Research*, *3*(9), 98-114.

Okoli, C., & Schabram, K. (2010). A guide to conducting a systematic literature review of information systems research.

Olagunju, M., & Utulu, S. (2021). Money Market Digitization Consequences on Financial Inclusion of Businesses at the Base of the Pyramid in Nigeria. *the digital distruption of financial services: international perspectives, Ewa Lechman & Adam Marszk (Eds.)*.

Oliver, E. (2010). Being Lecture Delivered at DBI/George Mason University Conferenceon Cyber Security holding. In: Department of Information Management Technology Federal University of ….

Osho, O., & Onoja, A. D. (2015). National Cyber Security Policy and Strategy of Nigeria: A Qualitative Analysis. *International Journal of Cyber Criminology*, *9*(1).

Parsons, S. (2020). The Duke of Cambridge visits the laboratory in Oxford where a potential vaccine has been produced. https://www.theguardian.com/world/2020/jul/17/russian-hackers-steal-coronavirus-vaccine-uk-minister-cyber-attack

Pattinson, M. R., Butavicius, M. A., Ciccarello, B., Lillie, M., Parsons, K., Calic, D., & McCormac, A. (2018). Adapting Cyber-Security Training to Your Employees. HAISA,

Pavol Zavarsky, C., & CISM, C. (2014). Step-by-step guidance on how to establish, implement and operate cybersecurity management system (ISMS).

Purohit, D. P., Siddiqui, N., Nandan, A., & Yadav, B. P. (2018). Hazard identification and risk assessment in construction industry. *International Journal of Applied Engineering Research*, *13*(10), 7639-7667.

Rashid, A., Danezis, G., Chivers, H., Lupu, E., Martin, A., Lewis, M., & Peersman, C. (2018). Scoping the cyber security body of knowledge. *IEEE Security & Privacy*, *16*(3), 96-102.

Ryder, R. D., & Madhavan, A. (2019). *Cyber Crisis Management: Overcoming the Challenges in Cyberspace*. Bloomsbury Publishing.

Sanoo, J. (2018). *Cyber Security Tutorials*. Retrieved 26/06/2020 from https://www.javatpoint.com/cyber-security-introduction

Saulawa, M. a. A., & Abubakar, M. (2014). Cybercrime in nigeria: An overview of cybercrime act 2013. *JL Pol'y & Globalization*, *32*, 23.

Sausalito, C. (2020). Cybercrime To Cost The World $10.5 Trillion Annually By 2025. *Cybercrime*. https://cybersecurityventures.com/hackerpocalypse-cybercrime-report-2016/#:~:text=Cybersecurity%20Ventures%20expects%20global%20cybercrime,%243%20trillion%20USD%20in%202015

Schneier, B. (2009). *Schneier on security*. John Wiley & Sons.

Singh, U. K., & Joshi, C. (2017). Information Security Risk Management Framework for University Computing Environment. *IJ Network Security*, *19*(5), 742-751.

Smith, W. (2019). A comprehensive cybersecurity defense framework for large organizations.

Sobers, R. (2021). *134 Cybersecurity Statistics and Trends for 2021*. https://www.varonis.com/blog/cybersecurity-statistics/

Soomro, T. R., & Hussain, M. (2019). Social media-related cybercrimes and techniques for their prevention. *Applied Computer Systems*, *24*(1), 9-17.

Stein, S. (2008). *ITU Global Cybersecurity Agenda (GCA) High-Level Experts Group (HLEG) Global strategic report*. ITU. https://www.itu.int/en/action/cybersecurity/Documents/gca-chairman-report.pdf.

Taylor, L. (2017). What is data justice? The case for connecting digital rights and freedoms globally. *Big Data & Society*, *4*(2), 2053951717736335.

Utulu, S., Sewchurran, K., & Dwolatzky, B. (2013). Systematic and Grounded Theory Literature Reviews of Software Process Improvement Phenomena: Implications for IS Research. Proceedings of the Informing Science and Information Technology Education Conference,

Utulu, S. C. A., & Ngwenyama, O. (2017). Model for constructing institutional framework for scientific knowledge management systems: Nigerian institutional repository innovation case applicable to developing countries. In *Catalyzing Development through ICT Adoption* (pp. 149-174). Springer.

Valeriano, B., & Maness, R. C. (2015). *Cyber war versus cyber realities: Cyber conflict in the international system*. Oxford University Press, USA.

Walker, A. (2020). UK '95% sure' Russian hackers tried to steal coronavirus vaccine research. https://www.theguardian.com/world/2020/jul/17/russian-hackers-steal-coronavirus-vaccine-uk-minister-cyber-attack

Webb, J., & Hume, D. (2018). Campus IoT collaboration and governance using the NIST cybersecurity framework.

Whitehead, G. (2020). *Investigation of factors influencing cybersecurity decision making in Irish SME's from a senior manager/owner perspective* Dublin, National College of Ireland].

Wolfswinkel, J. F., Furtmueller, E., & Wilderom, C. P. (2013). Using grounded theory as a method for rigorously reviewing literature. *European Journal of Information Systems*, 22(1), 45-55.

Xie, J. (2020). In Coronavirus Vaccine Hunt, a Race to Be First.